\documentclass{sig-alternate}
\usepackage{booktabs}
\newcommand{\ra}[1]{\renewcommand{\arraystretch}{#1}}

\begin{document}
%

\title{A new Definition and Classification of 
Physical Unclonable Functions}
%
%
%
%
%

\numberofauthors{2} 
%
\author{
%
%
\alignauthor
Rainer Plaga\\
       \affaddr{Federal Office for Information Security (BSI)}\\
       \affaddr{Godesberger Allee 185-189}\\
       \affaddr{Bonn, Germany}\\
       \email{rainer.plaga@bsi.bund.de}       
\alignauthor
Dominik Merli\\
       \affaddr{Fraunhofer Research Institution AISEC}\\
       \affaddr{Parkring 4}\\
       \affaddr{Garching (near Munich), Germany}\\
       \email{dominik.merli@aisec.fraunhofer.de}
}
\additionalauthors{Additional authors: John Smith (The Th{\o}rv{\"a}ld Group,
email: {\texttt{jsmith@affiliation.org}}) and Julius P.~Kumquat
(The Kumquat Consortium, email: {\texttt{jpkumquat@consortium.net}}).}
\date{30 July 1999}

\maketitle
\begin{abstract}
A new definition of ``Physical Unclonable Functions'' (PUFs), the first
one that fully captures its intuitive idea among experts, is presented.
A PUF is an  information-storage system with a security mechanism that is 
\\
1. meant to impede
the duplication of a precisely described storage-functionality 
in another, separate system and 
\\
2. remains effective against an attacker with temporary access to the 
whole original system.
\\
A novel classification scheme of the security objectives and mechanisms of PUFs
is proposed and its usefulness 
to aid future research and security evaluation 
is demonstrated.
One class of PUF security mechanisms that prevents an attacker to apply
all addresses at which secrets are stored in the information-storage
system, is shown to be closely analogous to cryptographic encryption.
Its development marks the dawn of a new fundamental primitive of hardware-security engineering:
cryptostorage.
These results firmly establish PUFs as a fundamental concept of hardware security.
\end{abstract}





\section{Introduction} 
\label{intro}
\emph{Physical Unclonable Functions} (PUFs)\cite{GassendSpr2002}
were originally
understood as ``hardware devices
that are hard to characterize and can be uniquely identified'' \cite{GassendSpr2002}.
They 
have developed into an important research topic in the field 
of hardware security within the past
decade. There have been a number of attempts (summarized in section
Section \ref{prev}) 
to precisely define what a member of the research community would 
intuitively call a PUF.
Searched for is a definition that contains a {\it complete and minimal}
list of the conditions that have to be fulfilled to identify
a device as a PUF.
No necessary condition must be absent. No superfluous condition, i.e. one
that PUFs can, but do not absolutely have to fulfill, must be present.
In section \ref{def} we present a novel definition of a PUF. We demonstrate
that it really fulfills, for the first time, both of the above demands.
\\
A novel classification scheme for the PUF security objectives and mechanisms
is presented in Section \ref{class}. Its usefulness is demonstrated
by two examples. In Section \ref{cs}
we show that the PUF security mechanism ``cryptostorage'' has
a fundamental importance similar to the one of cryptography. PUF research
is thus of a much more fundamental importance than hitherto realized.
We discuss how the new scheme motivates
qualitatively novel questions for further research in Section \ref{res}
and helps the certification
of PUFs in Section~\ref{cert} respectively.
Section~\ref{concl} concludes this paper.

\section{Definition of PUFs}

\subsection{Previous Proposals}
\label{prev}
In the following, we 
briefly sketch the development of the ideas on which we base
our proposal. Starting with an early paper by Gassend et al. \cite{GassendSpr2002},
most proposed definitions are based on the
concept of a physical challenge-response function
which can be evaluated in a reproducible manner. 
Until approx. 2010, most proposed definitions characterized
PUFs further by being ``hard to characterize''~\cite{GassendSpr2002},
``hard to predict''~\cite{GassendDCA2003},
``physically unclonable''~\cite{MaesDPP2010}
or ``tamper resistant''~\cite{SadeghiPRs2010}.
Such demands carry
the problem that devices which turn out to be predictable, clonable or
tamper vulnerable
are no longer PUFs according to the proposed 
definitions. However, e.g., in a certification process,
the concept of an ``insecure, broken PUF'' is clearly
necessary. Indeed, in practice, the
community does not really employ
such definitions because it universally continues to refer to insecure PUFs
that have been completely broken as ``PUFs''.
Armknecht et al. \cite{ArmknechtFSF2011} recognized this problem and defined PUFs as physical
functions PFs, i.e., they dropped the U for ``unclonable''. 
Such physical functions are digital memories
because the latter are physical devices that always have 
to realize a challenge-response mechanism: an
address is applied as a challenge and the memory content
is returned as a response.
The remaining problem is then: how to characterize the {\it specific
characteristic of PUFs} that sets it apart, e.g., from a standard
memory stick, which is, of course, not understood as a PUF in the community?
An alternative to identify it with absolute unclonability is to
identify it with certain features of the PUF architecture.
Several definitions \cite{RuhrmairSPM2010,MaitiSME2011,PlagaFDN2012} propose to identify the
specific characteristic properties that
we call below in Section \ref{class} ``complex-structure upon production''
and ``cryptostorage''.
We will discuss in Section \ref{sm} that these properties classify special
security mechanisms. But a demand for a
special security mechanism in its definition would render 
the PUF concept inflexible.
E.g. the information stored in a quantum token,
discussed in section \ref{classexist}, must be loaded
after its production - otherwise the storage is not secure.
A definition that demands a loading upon production would
rather arbitrarily excludes quantum tokens as PUFs.
\\
We  will explicitely identify
the {\it specific characteristic} of PFs, that we searched for in the 
previous paragraph in section \ref{mean}.

\subsection{Proposed Definition of PUFs and its \\ Meaning}
\label{def}
\subsubsection{Auxiliary and PUF Definition}
\label{deff}
\begin{figure}[!t]
\centering
\includegraphics[width=2.5in, height = 1.5 in]{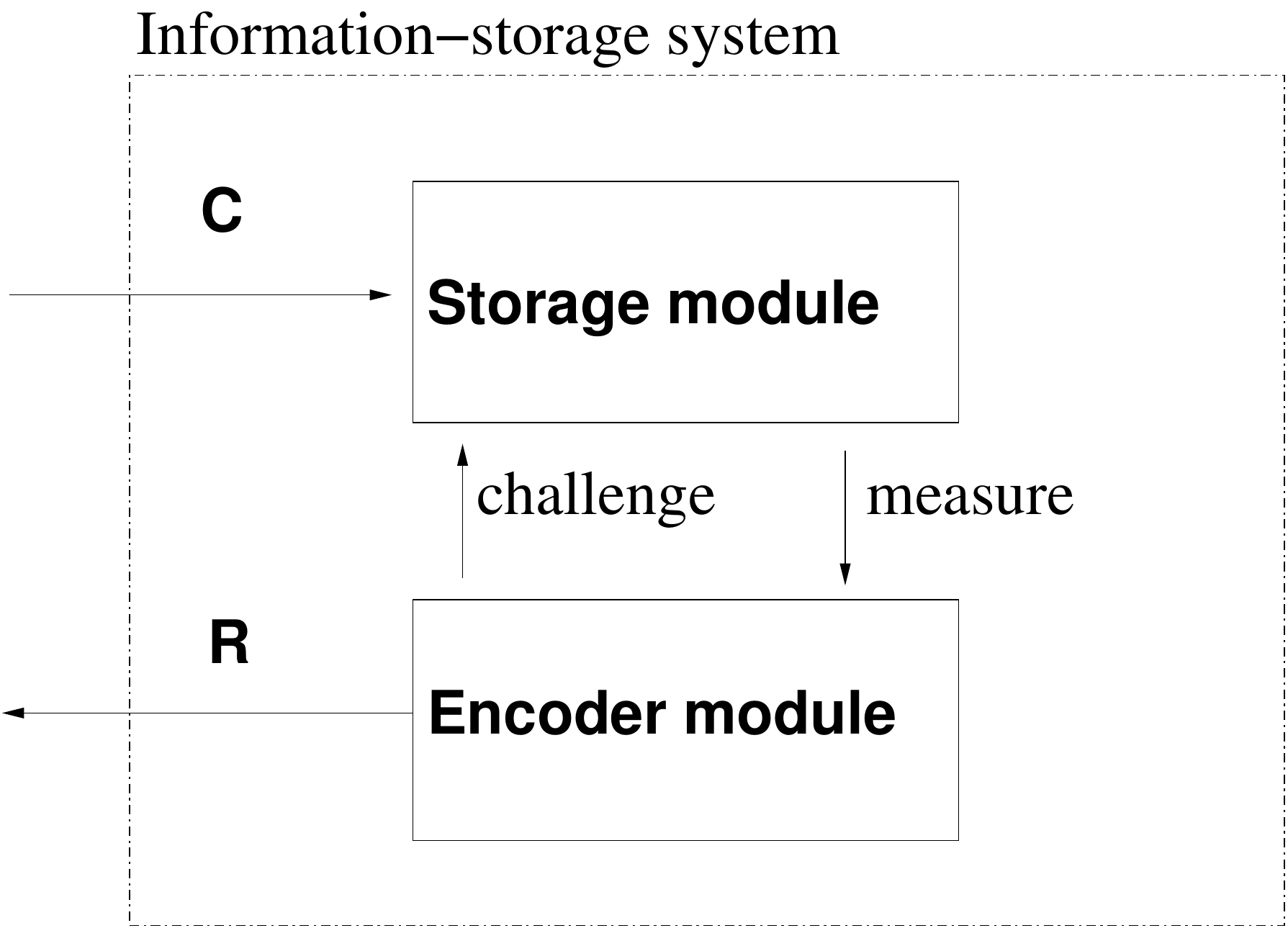}
\caption{\
Sketch of a physical information-storage system, according to our definition.
The dot-dashed outer box symbolizes the whole system which usually
is (but does not have to be) also a module, like the storage- and
encode-module always are.}
\label{explot}
\end{figure}

We first formulate an auxiliary definition:
\begin{itemize}
\item
{\bf Definition of a physical information-storage system (1)}
\\
{\it 
An information-storage system is a
set of at least two modules\footnote{A module is defined to be an artificial (the artificiality is 
necessary to delimit PUFs from
biometric systems, a closely related concept with which the PUF concept is often compared) 
physical system that is permanently physically
connected and serves a certain purpose. A physical system is defined to be a set of materials and fields that
is delimited from the rest of the world in a well defined manner.} that are conceptually
or physically permanently connected.
These modules (or parts of the module if the system is a single module) have the following purposes:
a ``storage module'' can be put into a physical state
H which is determined by the information in 
a challenge C. This storage module is measured in this state
and the measurement result is reproducibly encoded as the information of a response R
by another ``encoder module''
of the total system. The set of all Rs for all Cs is the information
stored in the system.}
\end{itemize}
This definition agrees with the intuitive
idea of a digital memory, and makes its relation
to a challenge-response function precise. 
The challenges are the addresses at which the physical information, the responses,
are stored.
An information-storage system is depicted in 
Figure \ref{explot}. The term ``reproducibly encoded'' demands
that a noise in the responses is bounded.

\begin{itemize}
\item
{\bf Definition of a PUF (2)}
\\
{\it 
A PUF is a physical information-storage system 
that is protected by a security mechanism 
which
\begin{enumerate}
  \item has the security objective to render it more
  difficult to duplicate a precisely described storage functionality
  of the system in another, separate system.

  \item is meant to remain effective against an active\footnote{Active means
that she can change the system rather than only
passively listen.} attacker with 
  temporary physical access
  to the whole system in its original form. 

\end{enumerate}
}
\end{itemize}

\subsubsection{Explanation of the PUF Definition}
\label{mean}
The definition does not demand that a security-mechanism actually does render the duplication more
difficult,
but only that an implemented mechanism is meant to do so.
This avoids the circularity problem with earlier PUF definitions discussed 
in Section \ref{intro} (``an insecure PUF is no PUF'').
As the following example shows
security primitives are often defined in an analogous manner:
\begin{itemize}
\item{\bf Definition of cryptographic encryption algorithm (3)}
\\
{\it A cryptographic encryption algorithm is a key-dependent mapping 
from a cleartext string of
bits to a cryptogram string of bits. Its
security objective is that an attacker finds no algorithm for an
inverse mapping (decryption)
of the cryptogram to the cleartext without the key.
The key is a string of bits on which the encryption algorithm depends.}
\end{itemize}
Caesar's cipher is trivial to break, but still a cryptographic encryption
algorithm because there is a specific key dependent mapping
with the said objective.
The precise storage functionality is the most characteristic feature of a PUF. 
It can go beyond the mere
storage and release of information, as in the case of public PUFs \cite{RuhrmairSSP2009,BeckmannHpc2009},
where a release of the response within a certain time frame is required.
Even if the functionality is merely the storage
and release of physical information,
the objective can be to prevent duplication
to different degrees as in the 
previous proposals of
a ``physical'' and ``mathematical'' duplication \cite{MaesDPP2010}.
We will come back to the classification of PUF security
objectives in Section \ref{so}.
We define an ``information-storage system'' as the system
that actually stores the information (rather than auxiliary
systems for packaging, energy supply etc.).
Our definition of a PUF requires that 
the security mechanism is based on the system's
storage-mechanism, because otherwise it would not
remain effective against an attacker with full access to the
original system. 
This makes the required security- and information-storage mechanism
indivisible, i.e., without the latter, the former cannot be realized.
Devices in which information-storage and security mechanism are
separable are no PUFs. 
This indivisibility, let us call it ``security-memory boundedness'', is the
{\it specific characteristic} 
of a PUF that was searched for in Section \ref{prev}.
\\
Another angle on the motivation for the second 
condition of the PUF definition will be discussed in Section \ref{delim}.
\\
A main result of this paper is that the PUF definition list is complete
and minimal with the proposed conditions. 
No other of the myriad other conditions that PUFs may well fulfill, e.g.,
that the storage mechanism is noisy, that the security mechanism works
without energy supply, etc., has to be fulfilled to characterize
a module as a PUF. Summarizing we state:
\\
{\bf A module is a PUF if
and only if it fulfills both conditions of the definition (2).}

\subsubsection{Exemplary Analysis: Arbiter PUF \\ Characterized as a PUF}

As an exemplary case, we discuss how the arbiter PUF \cite{SuhPUF2007}
is characterized as a PUF in the sense of our definition.
The arbiter PUF has a storage module in the form of a chain of multiplexers
that are programmed by a certain number of inputs which act as a challenge. 
The encoder module is a latch that
determines which of two delay paths - that are configured (with other
words ``put in a physical state'') by the challenge - produced
a longer delay for two test signals. The arbiter PUF
is thus an information-storage system in our sense. 
The security
mechanism is described by Suh and Devadas
as: 
``a set of challenges (is mapped)
to a set of responses based on an intractably complex physical system.''
The unclonable storage-functionality is the release of a response after
a challenge is supplied.
The authors make clear that this must be understood as an objective rather
than a property because
``to prevent model-building attacks'' a further
development of the PUF architecture might
be needed. Thus the first condition of our definition of
a PUF is fulfilled. The ``intractably complex structure'' upon production
is meant to protect against an attacker with temporary access to the storage system.
Therefore, the second condition of
our definition (2) is fulfilled, too.

\subsubsection{Delimitation of PUFs from other Concepts}
\label{delim}
Conventional secure memories, as typically realized, e.g., in smartcards
as conventional digital memories with some passive or active protection shields
to guarantee ``tamper resistance'',
are no PUFs because they are not security-memory bounded.
A conventionally secured memory has been called ``controlled'' by
Gassend et al.\cite{GassendCPR2008}
\footnote{Gassend et al. call a PUF controlled ``if it can only
be accessed via an algorithm that is linked to the PUF
in an inseparable way (i.e. any attempt
to circumvent the algorithm will lead to the destruction 
of the PUF.'' If we replace ``PUF'' by ``information-storage
system'' and interpret ``algorithm'' as ``hardware-interface that implements
the system's intended functionality'', this is a definition
of the hitherto conventional manner to protect an information-storage system
against duplication.}.
The qualitative novelty of PUFs, characterized by the second
condition in the definition (2), is that their security
mechanisms go beyond such conventional (access) control. 
\\
Unique physical labels as they are used, e.g., on banknotes or
drug packages against counterfeiting are no PUFs,
not conceptually  permanently connected to an encoder module
as required by our definition (2).
``Physically obfuscated keys'' \cite{RuehrmairFoP2009} with a read-out
mechanism for the key are PUFs because their obfuscation 
through a non-standard storage mechanism is security-memory bound.

\section{PUF Classification}
\label{class}

\subsection{Classification Criteria}

\begin{figure}[!t]
\centering
\includegraphics[angle=0,width=2.5in]{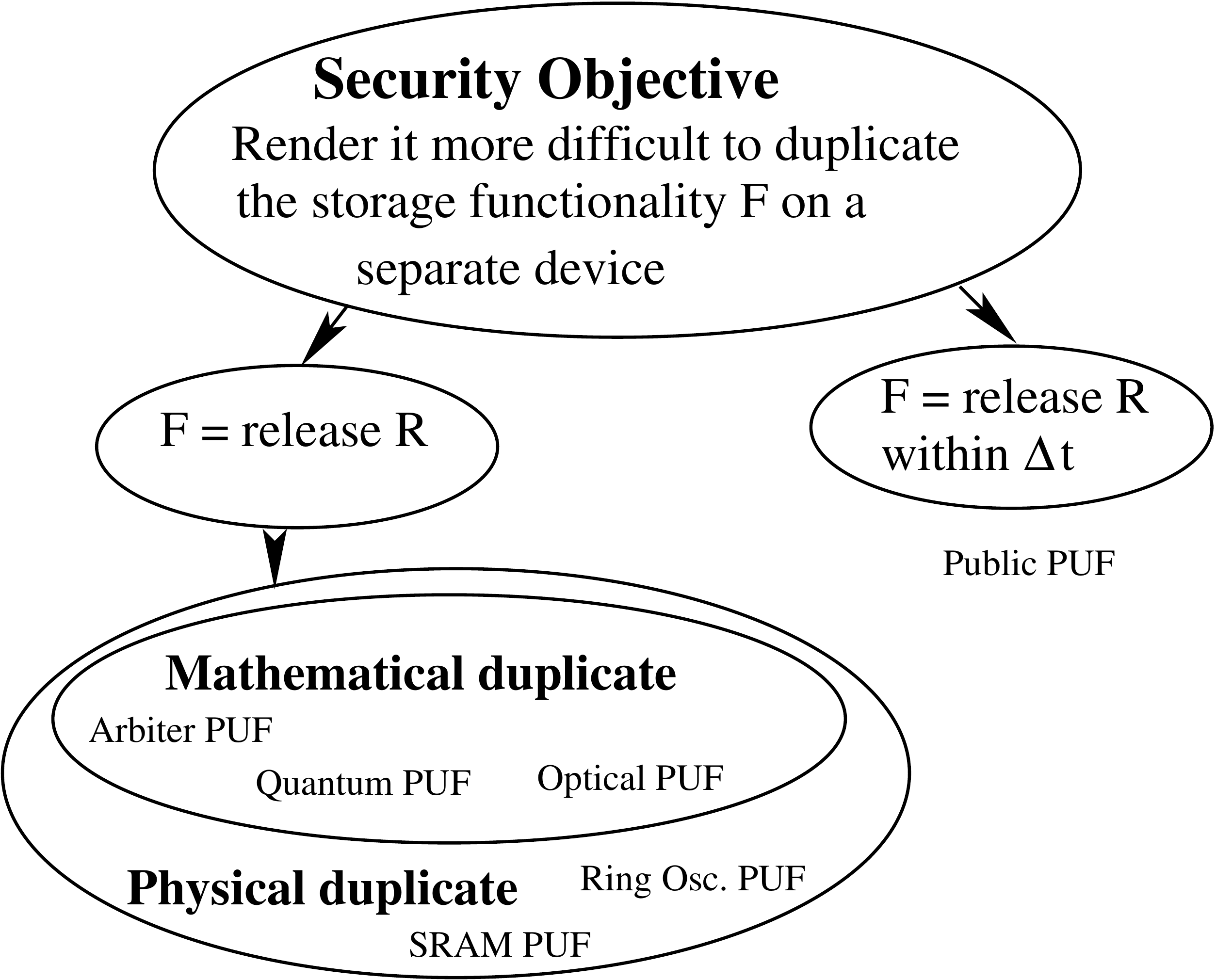}
\caption{
Classification of the PUFs' security objectives; some existing
PUFs are assigned to these classes.} 
\label{obj}
\end{figure}

\subsubsection{Security Objectives}
\label{so}
The most basic classification is by security objectives.
According
to the first condition in
definition (2)
the objectives are characterized by the precise nature of the
storage functionalities and the kind of duplication to be prevented.
The simplest storage functionalities are:
\\ 
\noindent\hspace*{6 mm}%
{\it S1. To store physical information and
to release it upon some challenge which can be:} 
\\
\noindent\hspace*{12 mm}%
{\it S1.a. a simple trigger or}
\\
\noindent\hspace*{12 mm}%
{\it S1.b. a sophisticated address, chosen by the user.}
\\
An example for a more sophisticated storage functionality is:
\\
\noindent\hspace*{6 mm}%
{\it S2. the release of the
stored information within a certain time interval $\Delta$t.}
\\
There are two widely used
meanings of the term ``duplicate''\cite{MaesDPP2010}.
\\
\noindent\hspace*{6 mm}%
{\it D1. ``Physical duplication'' is to prevent a duplication in physical detail. I.e.
the separate duplicated system 
either follows the specifications for the construction of the
original, or at least returns physically identical responses.
\\
\noindent\hspace*{6 mm}%
D2. ``Mathematical duplication'' only
requires that the duplication system returns responses
with the same information content 
as the ones of the original module to all challenges. Its physical structure is otherwise
not constrained.}
\\
D2 is a more ambitious objective than D1 because it requires to 
absolutely prevent the readout of the
stored information at the challenge addresses, also against 
sophisticated modeling attacks\cite{RuehrmairACM2010}.
As soon as this information is read out, it can
be stored in another memory under the same address thus realizing a mathematical clone.
D1 can still be attainable when the stored information is known to the attacker.
\\
Fig. \ref{obj} illustrates the storage-functionality classes and gives examples
of PUFs within them. 
\\
The security objectives are determined by the security architecture that uses
the PUF rather than the architecture of the PUF itself. In Section \ref{classexist},
we will discuss classification examples for
some PUFs including one that can have both objective D1 or D2 in different contexts.
\newpage
\subsubsection{Security Mechanisms}
\label{sm}
\begin{figure}[!t]
\centering
\includegraphics[angle=0,width=2.5 in,height=2.0 in]{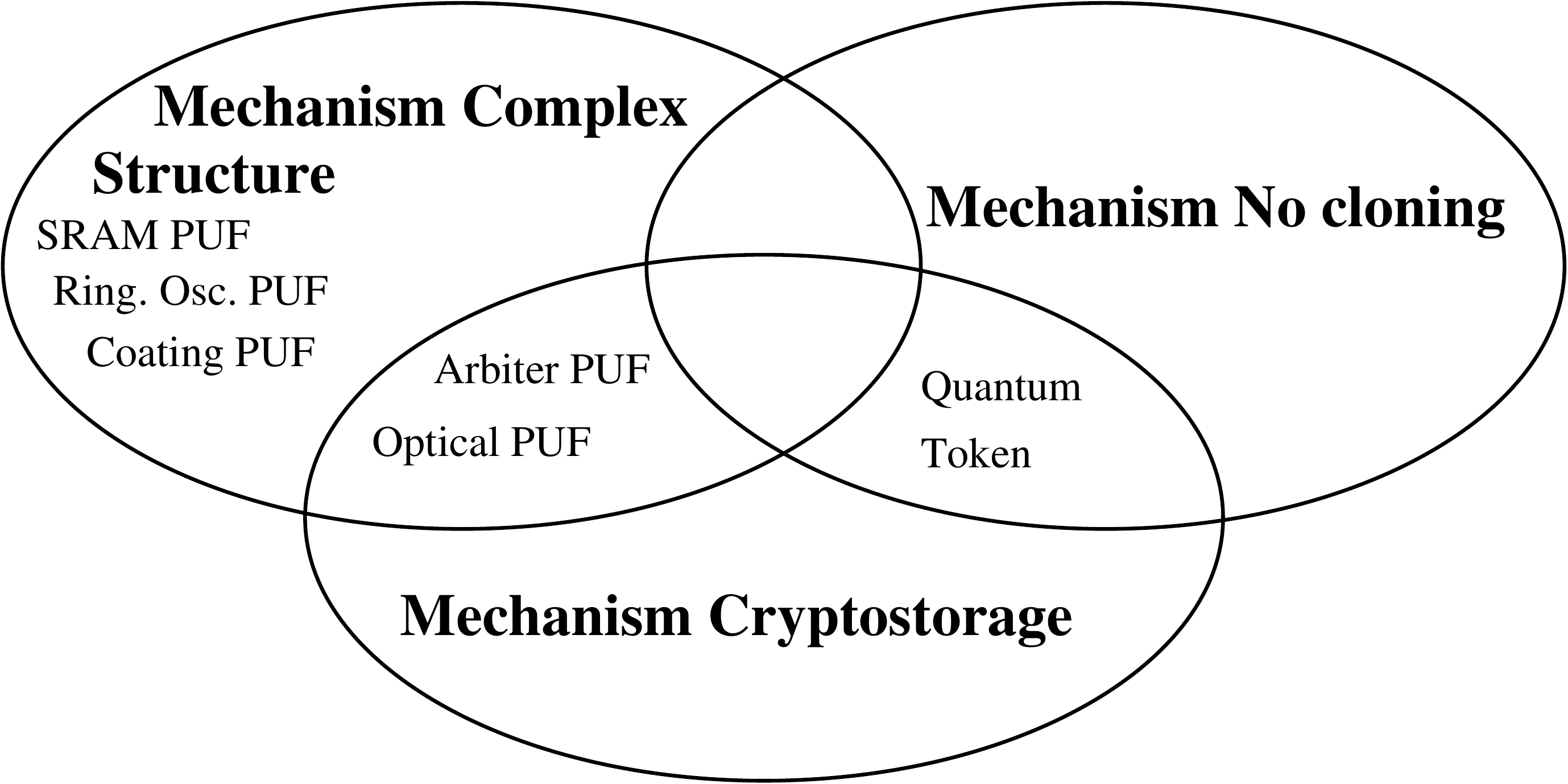}
\caption{
The PUF security mechanisms and PUFs that
are based on them. 
See Section \ref{sm} for further explanation.}
\label{mechplot}
\end{figure}

Three principal classes of security mechanisms for
PUFs have been proposed up to now.
\\
\noindent\hspace*{6 mm}%
{\it Security mechanism 1 ``Complex Structure''(CS). A complex structure of the
information-storage module
prevents analysis and/or
reproduction.}
\\
Most PUFs that were proposed up to now rely on this principle. Usually the following principle
is exploited: it is easier to create a complex random structure, than to analyze and duplicate it. 
Such PUFs are produced in a process that
creates some complex structure in the storage module
from which random responses can be derived by
the encode module\cite{MaesCHES2012}. 
We classify this
security mechanism as PUFs with ``Complex-Structure upon Production'' (CSP) property.
\\
\noindent\hspace*{6 mm}%
{\it Security mechanism 2 ``No Cloning''(NC). A principle
of physics forbids the duplication of
the storage module.}
\\
This is a security mechanism that prevents
the duplication of the storage unit because of some
fundamental principle of physics. An example are
physical quantum systems in 
a state unknown to the attacker, that cannot
be cloned due to the ``no cloning principle''\cite{PastawskiUNQ2011}.
\\
\noindent\hspace*{6 mm}%
{\it Security Mechanism 3 ``Cryptostorage''.
Let us call a subset of
all possible challenges the
set of secret challenges (s-challenges).
The secret responses (s-responses) to be protected are defined
as the responses to the s-challenges.
The security mechanism prevents that
the attacker can apply all or most s-challenges.}
\\ 
If the attacker cannot apply an s-challenge in principle 
she will not be able
to get the corresponding secret (response), i.e. cryptostorage protects
the physical stored information against read out.
Cryptostorage is closely analogous to cryptographic encryption.
This parallel will be further
developed in Section \ref{cs}.
\\
Two general architectures to realize cryptostorage 
in practice have been proposed
\cite{PlagaFDN2012}.
\\
{\it Security mechanism 3a. ``Minimum Readout Time (MRT)''.
The number of challenge-response pairs is
chosen very large. Then, if
there is a sufficiently long minimum readout time for
one challenge-response pair, the attacker
cannot apply the s-challenges within a reasonable
time period if the set of s-challenges is unknown to her.}
\\
PUFs with a MRT-mechanism are a.k.a. as ``strong'' PUFs \cite{RuehrmairFoP2009}. 
\\
{\it Security mechanism 3b. ``Erasure Upon Readout (EUR)''.
If a non s-challenge is applied the
s-response corresponding to the s-challenge is erased. Therefore the attacker
cannot apply the s-challenge if it is not known
to her, because an attempt to do so, erases most of the secret
information in general.}
\\
Fig. \ref{mechplot} illustrates which
PUF constructions are protected
by these mechanisms.

\subsection{Exemplary Classification of some Existing PUFs}
\label{classexist}
The ring-oscillator PUF \cite{SuhPUF2007} and the SRAM PUF \cite{GuajardoFIP2007}
exploit variations
in ring oscillator frequencies or memory cell balances respectively, to generate secret bits
and store the information in a currently unconventional manner. The security
mechanism is to generate a random value during fabrication i.e. it is a 
CSP mechanism. 
The security objective of the
ring-oscillator PUF \cite{SuhPUF2007} and the SRAM PUF \cite{GuajardoFIP2007} 
(a.k.a ``weak PUFs'' \cite{RuehrmairFoP2009})
can be twofold.
One possible aim is to prevent the physical duplication (security objective D1 in Section \ref{so}). 
This security objective will typically arise in 
a context in which one aims to make it difficult for the producer of the PUF to produce
physical duplicates.
The other possible aim is to render the mathematical duplication (security objective D2 in Section \ref{so})
more difficult by making a
readout in the switched-off state more difficult. 
This second aim cannot reasonably be to prevent
the read-out of the stored information altogether
because no security mechanism of SRAM and ring-oscillators prevents
such readout.
\\
The arbiter PUF \cite{LimEsk2005}
uses intrinsic delay variations in microchips and was proposed 
to be used with an s-challenge that is much smaller
than the set of all challenges. It thus
aims to prevent the mathematical duplication (security objective D2, in Section \ref{so})
with the help of both the CSP mechanism and cryptostorage with the MRT mechanism.
For quantum-token based PUFs \cite{PastawskiUNQ2011}, the security
objective D2 is reached via
proving that the s-challenge is
necessary to extract the previously stored information. 
Therefore, this PUF prevents mathematical duplication with the help of the NC mechanism and  
cryptostorage with the EUR mechanism. 
\\
A qualitatively special authentication
functionality is realized with the 
``public PUF'' \cite{RuhrmairSSP2009,BeckmannHpc2009} idea, which
proposes PUFs that can prove their authenticity even though
their challenge-response pairs are public. A public PUF not
only stores and protects information, but is also required
to respond to the challenge within a
certain minimum time period. It has the objective to prevent 
the construction of a duplicate that releases a response
within a certain time period, even though a module that
releases it within a much longer period must be possible.

\section{A New Security Primitive: \\ Cryptostorage}
\label{cs}
\subsection{Description of the Primitive}
The security mechanism of ``cryptostorage'' first
introduced in Section \ref{sm} is of fundamental
importance. It can be characterized in the following manner:
\begin{itemize}
\item{\bf Definition of the cryptostorage mechanism (4)}
\\
{\it 
A clearstring is a string of bits physically stored in
an information-storage system with no security mechanism.
A cryptostorage mechanism is a s-challenge-dependent mapping 
from a clearstring to the storage module of a PUF. Its
security objective is that an attacker finds no mechanism
for the extraction
of the clearstring from the PUF without the set of s-challenges.
The s-challenge is a string of bits that designates a subset
of the address space of the information-storage system.}
\end{itemize}

When one compares this definition with the
one of a cryptographic encryption algorithm (3),
the close analogy between the two is obvious:
the s-challenge is equivalent to the key, the storage
system of the PUF to the cryptogram and the 
clearstring to the cleartext.
If cryptographic encryption is the science of concealed (crypto) writing (graphein),
research and development on PUFs can be identified as the science of concealed storage,
or ``cryptostorage''. 
Cryptostorage
protects the physical information\footnote{Physical information is a physical
system that has a certain computable relation to the outside world in space an time.} itself.
Cryptographic encryption protects the meaning of 
the information\footnote{The meaning of information is a description of the relation between
physical information and outside world.} in the cryptogram. 
The close parallel between cryptographic encryption and cryptostorage
is illustrated in Table \ref{cana}.
\\
In cryptographic encryption, principles and
comprehensible systematic arguments (not necessarily proofs)
based on mathematics and informatics 
ensure that the cleartext cannot be extracted
without the key. In an analogous manner, in cryptostorage,
principles based on physics and electrical engineering must
ensure that the stored information cannot
be extracted without being in possession of the s-challenges.
The analog of
cryptanalytic analyses of cryptographic encryption are, e.g.,
attacks using various forms of sophisticated learning programs\cite{RuehrmairACM2010}.
\\
The specific value of cryptostorage for embedded security is the systematic practical
and theoretical development
of information-storage systems that remain secure when the attacker can directly and completely 
access them, i.e. uses their regular read-out system.
It will be necessary to develop the primitive of cryptostorage in a systematic manner because it will
always remain impossible in principle to completely deny access to embedded storage systems.
The major aim of cryptostorage will be to develop PUFs with a comprehensible, well understood
security level based on sound principles of engineering, physics and mathematics.

\begin{table}
\ra{1.3}
\begin{tabular}{lll}
\toprule
 & {\bf Cryptographic encryption} & {\bf Cryptostorage} \\
\midrule
& 1. meaning of Information & physical Information  \\
& 2. cryptogram & PUF storage module \\
& 3. crypto-algorithm & cryptostorage mechanism \\
& 4. inversion of cryptogram & reproduction of PUF \\
& 5. cryptographic key & s-challenge \\
& 6. encrypt with key & store at s-challenge adr. \\
& 7. decrypt with key & apply s-chall. \& measure \\
\bottomrule
\end{tabular}
\caption{
A listing of corresponding concepts of cryptographic encryption and cryptostorage.
1. What is protected? 2. Where is protected information? 3. What is the security
mechanism that 4. prevents what?
}
\label{cana}
\end{table}

\section{Future PUF Research}
\label{res}

The two most basic question are:
\\
1. What new security objectives for PUFs could there be?
Which PUF storage-functionalities, besides the 
simple, addressed and timed release 
of information could be useful?
\\
2. What new security mechanisms, besides the CS, NC and 
cryptostorage classes could there be?
\\
The cryptostorage security primitive is bound to play
a similarly important role in embedded security as
cryptographic encryption does. 
Many proposed PUFs are based on it, but the security
of these proposals still need further study.
The foundations of cryptostorage need to be developed systematically:
Which further standard design principles for physical memories could be given up, to
make them security-memory bounded?
How can EUR PUFs be constructed which are not based on quantum-storage?
How can it be ensured that MRT PUFs with $N$ secret elements
require on the order of $2^N$ challenge-response pairs to extract
all contained information even under sophisticated learning attacks?
Another important research topic are algorithms and protocols to support PUF-based applications.
\\
If a new ``PUF protocol'', e.g., for authentication is proposed, it should be communicated
which of its features are PUF specific, and why.
Otherwise, it would be a general authentication protocol, i.e., its novelty
cannot lie in its applicability to PUFs.

\section{Certification of PUFs}
\label{cert}
A first step in a PUF security evaluation must be to ascertain that
a hardware element within a target of evaluation (TOE) is really a PUF according to our proposed definition.
\\
The next step will be to identify the PUF's security objectives
within the TOE's architecture and the security mechanisms that
fulfill these objectives.
The central task of a PUF certification will be to theoretically
model and verify the PUF's security mechanism in simulation and practice.
Possible effects of the PUF's security mechanism on other conventional, non-PUF
mechanism that secure the protected memory also need to be considered.
\\
Classifying  the PUF according to the
characteristics in Section \ref{class} will be useful to lay out the scope
of the evaluation. Precisely defined security objectives clearly determine
the tasks of the security mechanisms.
If, e.g., the PUF has
an internally generated random value as response (CSP mechanism), the machine
producing it assumes the role of a Random-Number Generator (RNG). It will
then be necessary to evaluate the manufacturing system, e.g., according to
the existing guidelines for the certification of RNGs \cite{KillmannAIS312013}.
\\
PUFs can be valuable because they reach new levels of security, but 
also because they allow cheaper solutions than conventional approaches.
Therefore, security evaluations of PUFs must not just
address whether a PUF can be reproduced but also at which effort.

\section{Conclusion}
\label{concl} We proposed to
define PUFs as physical memories designed with the objective to
protect a well defined storage functionality against duplication with a mechanism that
is inseparable from the information-storage mechanism.
PUFs in this sense seem to completely characterize what
the community always meant by this concept.

We welcome challenges to this claim.
\\
Our definition does not restrict PUF architectures in any way, but provides
a clear separation to other conventional security modules. Conventional
secure memories in which information-storage and security mechanism are
separable, i.e. protect against access to the informations-storage system, are no PUFs. 
In principle, their security mechanism can also protect entities 
different from associated memories. Therefore, a classification
of such architectures as two independent security and storage
architectures is more appropriate.
\\
We argue that PUF research and development is
a much more fundamental and important field than 
previously thought. 
In a sense, it is the discipline of
the secure storage of physical information itself, just
as cryptography is the field of secure writing of
the meaning of information.
\\~\\

\section*{Acknowledgment}
We thank
Ralph Breithaupt, Ute Gebhardt, Dieter Schuster and Markus Ullmann for helpful discussions and
criticisms on previous drafts of this manuscript. 
We further thank three anonymous referees for very helpful critical remarks.

%
\bibliographystyle{abbrv}
\bibliography{PUF-host2b}  

\begin{thebibliography}{10}
\providecommand{\url}[1]{#1}
\csname url@samestyle\endcsname
\providecommand{\newblock}{\relax}
\providecommand{\bibinfo}[2]{#2}
\providecommand{\BIBentrySTDinterwordspacing}{\spaceskip=0pt\relax}
\providecommand{\BIBentryALTinterwordstretchfactor}{4}
\providecommand{\BIBentryALTinterwordspacing}{\spaceskip=\fontdimen2\font plus
\BIBentryALTinterwordstretchfactor\fontdimen3\font minus
  \fontdimen4\font\relax}
\providecommand{\BIBforeignlanguage}[2]{{%
\expandafter\ifx\csname l@#1\endcsname\relax
\typeout{** WARNING: IEEEtran.bst: No hyphenation pattern has been}%
\typeout{** loaded for the language `#1'. Using the pattern for}%
\typeout{** the default language instead.}%
\else
\language=\csname l@#1\endcsname
\fi
#2}}
\providecommand{\BIBdecl}{\relax}
\BIBdecl

\bibitem{GassendSpr2002}
B.~Gassend, D.~Clarke, M.~van Dijk, and S.~Devadas, ``Silicon physical random
  functions,'' in \emph{CCS '02: Proceedings of the 9th ACM conference on
  Computer and communications security}.\hskip 1em plus 0.5em minus 0.4em\relax
  New York, NY, USA: ACM, 2002, pp. 148--160.

\bibitem{GassendDCA2003}
------, ``Delay-based circuit authentication and applications,'' in
  \emph{Symposium on Applied Computing (SAC)}, 2003.

\bibitem{MaesDPP2010}
R.~Maes and I.~Verbauwhede, ``A discussion on the properties of physically
  unclonable functions,'' TRUST 2010 Workshop on Security Hardware, Berlin, DE,
  2010.

\bibitem{SadeghiPRs2010}
A.-R. Sadeghi, I.~Visconti, and C.~Wachsmann, ``{PUF}-enhanced rfid security
  and privacy,'' in \emph{Workshop on Secure Component and System
  Identification (SECSI)}, 2010.

\bibitem{ArmknechtFSF2011}
\BIBentryALTinterwordspacing
F.~Armknecht, R.~Maes, A.-R. Sadeghi, F.-X. Standaert, and C.~Wachsmann, ``A
  formalization of the security features of physical functions,'' in
  \emph{Proceedings of the 2011 IEEE Symposium on Security and Privacy}, ser.
  SP '11.\hskip 1em plus 0.5em minus 0.4em\relax Washington, DC, USA: IEEE
  Computer Society, 2011, pp. 397--412. [Online]. Available:
  \url{http://dx.doi.org/10.1109/SP.2011.10}
\BIBentrySTDinterwordspacing

\bibitem{RuhrmairSPM2010}
U.~R\"{u}hrmair, H.~Busch, and S.~Katzenbeisser, ``Strong {PUFs}: Models,
  constructions, and security proofs,'' in \emph{Towards Hardware-Intrinsic
  Security}, A.-R. Sadeghi and D.~Naccache, Eds.\hskip 1em plus 0.5em minus
  0.4em\relax Springer, 2010, pp. 79--96.

\bibitem{MaitiSME2011}
A.~Maiti, V.~Gunreddy, and P.~Schaumont, ``A systematic method to evaluate and
  compare the performance of physical unclonable functions,'' Cryptology ePrint
  Archive, Report 2011/657, 2011, http://eprint.iacr.org/2011/657.

\bibitem{PlagaFDN2012}
R.~Plaga and F.~Koob, ``A formal definition and a new security mechanism of
  physical unclonable functions,'' in \emph{Proceedings of the 16th
  international GI/ITG conference on Measurement, Modelling, and Evaluation of
  Computing Systems and Dependability and Fault Tolerance}, ser.
  MMB'12/DFT'12.\hskip 1em plus 0.5em minus 0.4em\relax Berlin, Heidelberg:
  Springer-Verlag, 2012, pp. 288--301.

\bibitem{RuhrmairSSP2009}
U.~R\"{u}hrmair, ``Simpl systems: On a public key variant of physical
  unclonable functions,'' Cryptology ePrint Archive, International Association
  for Cryptologic Research, Tech. Rep., 2009.

\bibitem{BeckmannHpc2009}
N.~Beckmann and M.~Potkonjak, ``Hardware-based public-key cryptography with
  public physically unclonable functions,'' in \emph{Information Hiding}.\hskip
  1em plus 0.5em minus 0.4em\relax Springer, 2009, pp. 206--220.

\bibitem{SuhPUF2007}
G.~E. Suh and S.~Devadas, ``Physical unclonable functions for device
  authentication and secret key generation,'' \emph{Design Automation
  Conference, 2007. DAC '07. 44th ACM/IEEE}, pp. 9--14, 2007.

\bibitem{GassendCPR2008}
B.~Gassend, M.~van Dijk, D.~Clarke, E.~Torlak, P.~Tuyls, and S.~Devadas,
  ``Controlled physical random functions and applications,'' \emph{ACM
  Transactions on Information and System Security}, vol.~10, no.~4, Jan. 2008.

\bibitem{RuehrmairFoP2009}
U.~R\"{u}hrmair, J.~S\"{o}lter, and F.~Sehnke, ``On the foundations of physical
  unclonable functions,'' Cryptology ePrint Archive, Report 2009/277, 2009,
  http://eprint.iacr.org/.

\bibitem{MaesCHES2012}
R.~Maes, A.~V. Herrewege, and I.~Verbauwhede, ``Pufky: A fully functional
  puf-based cryptographic key generator,'' in \emph{CHES}, vol. 7428.\hskip 1em
  plus 0.5em minus 0.4em\relax Springer, 2012, pp. 302--319.

\bibitem{PastawskiUNQ2011}
F.~Pastawski, N.~Y. Yao, L.~Jiang, M.~D. Lukin, and J.~I. Cirac, ``Unforgeable
  noise-tolerant quantum tokens,'' 2011.

\bibitem{GuajardoFIP2007}
J.~Guajardo, S.~S. Kumar, G.~J. Schrijen, and P.~Tuyls, ``{FPGA} intrinsic
  {PUF}s and their use for {IP} protection,'' in \emph{9th International
  Workshop of Cryptographic Hardware and Embedded Systems, CHES'07}, 2007, pp.
  63--80.

\bibitem{LimEsk2005}
D.~Lim, J.~W. Lee, B.~Gassend, G.~E. Suh, M.~van Dijk, and S.~Devadas,
  ``Extracting secret keys from integrated circuits,'' \emph{Very Large Scale
  Integration (VLSI) Systems, IEEE Transactions on}, vol.~13, no.~10, pp.
  1200--1205, December 2005.

\bibitem{RuehrmairACM2010}
U.~R\"{u}hrmair, F.~Sehnke, J.~S\"{o}lter, G.~Dror, S.~Devadas, and
  J.~Schmidhuber, ``Modeling attacks on physical unclonable functions,'' in
  \emph{Proceedings of the 17th ACM conference on Computer and communications
  security}, ser. CCS '10.\hskip 1em plus 0.5em minus 0.4em\relax New York, NY,
  USA: ACM, 2010, pp. 237--249.

\bibitem{KillmannAIS312013}
W.~S. Wolfgang~Killmann, ``Ais 31 v3 - functionality classes and evaluation
  methodology for random number generators,'' Federal Office for Information
  Security (BSI), Germany, 2013,
  https://www.bsi.bund.de/SharedDocs/Downloads/DE/BSI/Zertifizierung
  /Interpretationen/AIS\_31\_pdf.pdf.

\end{thebibliography}
%
%


\end{document}